\documentclass[12pt, journal,draftcls,letterpaper,onecolumn]{IEEEtran}

\usepackage{amssymb}
\usepackage{bbm}
\usepackage{pdfsync}

\usepackage{cite}

\ifCLASSINFOpdf
   \usepackage[pdftex]{graphicx}
   \graphicspath{{./EPS_FIGS/}}
   
   \DeclareGraphicsExtensions{.pdf}
   
\else
\fi
\usepackage[cmex10]{amsmath}
\usepackage{amsfonts}

\begin{document}
%
\title{Analog Baseband Cancellation for Full-Duplex: An Experiment Driven Analysis}
%
%
%


\author{\IEEEauthorblockN{Brett~Kaufman, \IEEEmembership{Student~Member,~IEEE,} Jorma~Lilleberg, \IEEEmembership{Senior~Member,~IEEE,}
\\and~Behnaam~Aazhang, \IEEEmembership{Fellow,~IEEE}}

\thanks{B. Kaufman, J. Lilleberg, and B. Aazhang are jointly with the Center for Multimedia Communication at Rice University and the Centre for Wireless Communication at the University of Oulu, Finland.  J. Lilleberg is also with Renesas Mobile in Oulu, Finland.  This work is funded in part by NSF, the Academy of Finland through the Co-Op grant, and by Renesas through a research contract.}}
\maketitle

\begin{abstract}
Recent wireless testbed implementations have proven that full-duplex communication is in fact possible and can outperform half-duplex systems.  Many of these implementations modify existing half-duplex systems to operate in full-duplex.  To realize the full potential of full-duplex, radios need to be designed with self-interference in mind.  In our work, we use a novel patch antenna prototype in an experimental setup to characterize the self-interference channel between transmit and receive radios.  We derive an equivalent analytical baseband model and propose analog baseband cancellation techniques to complement the RF cancellation provided by the patch antenna prototype.   

Our results show that a wide bandwidth, moderate isolation scheme achieves up to 2.4 bps/Hz higher achievable rate than a narrow bandwidth, high isolation scheme.   Furthermore, the analog baseband cancellation yields a $10^1 - 10^4$ improvement in BER over RF only cancellation.   

\end{abstract}



%
\IEEEpeerreviewmaketitle

\section{Introduction}
\begin{figure}[htp] 
\begin{center} 
  \includegraphics[width=0.493\textwidth]{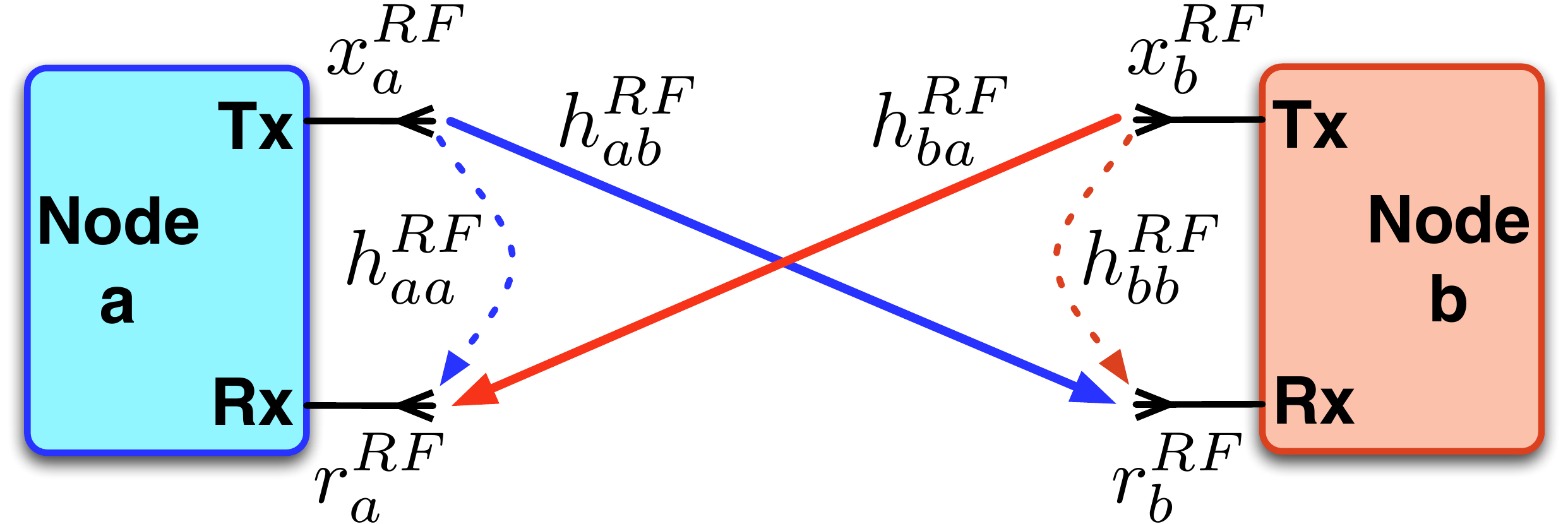}
  \caption[fig:system]{Two-user full-duplex link showing the self-interference channels $h_{aa}^{RF}$ and $h_{bb}^{RF}$ with dashed arrows and data channels $h_{ab}^{RF}$ and $h_{ba}^{RF}$ with solid arrows.} 
  \label{fig:network}
\end{center}    
\end{figure}

In the constant pursuit of increased spectral efficiency and faster data rates, full-duplex communication \cite{2010_Melissa_Asilomar,2010_Stanford_Mobicom} has emerged as a highly promising technique.  Full-duplex communication occurs when a node simultaneously transmits and receives information on the same frequency band.  In the context of a wireless point-to-point link, shown in Fig.~\ref{fig:network}, full-duplex enables two nodes to communicate over a bidirectional link using the same temporal and spectral resources.  This is in stark contrast to the half-duplex mode in which most currently deployed wireless systems operate in.  Half-duplex communication modes utilizes either time-division or frequency-division to allocate each node roughly half of the available resources.  

The major limiting factor in realizing full-duplex communication is the strong self-interference originating from a node's own transmit antenna.  Due to the relatively close proximity between a node's transmit and receive antennas, the strength of the self-interference can be significantly larger (50-100 dB) than the signal of interest sent by another node.  Until very recently, it was believed that the self-interference signal was too strong for full-duplex to even be feasible.  Several notable implementations \cite{2011_Evan_Asilomar, 2012_Melissa_TWC, 2011_Stanford_Mobicom} were demonstrated showing that varying amounts of the self-interference could in fact be canceled and provide positive gains over half-duplex mode.
  
In order to maximize the gains of full-duplex, the self-interference needs to be cancelled in its entirety. Numerous research directions have been proposed with this end goal in mind.  Below is a brief survey of notable full-duplex research directions with special emphasis on self-interference cancellation.

\subsection{Related Work}
\label{subsec:related_work}
Self-interference cancellation can be broadly separated into two main techniques, \emph{Passive Suppression} and \emph{Active Cancellation}.  Passive techniques are agnostic to the specific signal that a node is transmitting and attempts to increase the attenuation between the transmit and receive signals.  
A majority of the passive techniques rely on antenna design and/or placement to suppress the self-interference.  Experiments in \cite{2013_Evan_TWC_passive} used foam absorption material to isolate the transmit and receive antennas in addition to polarized antennas to orthogonalize the transmit and receive data streams.  Implementations in \cite{2011_Antenna_cancel,2012_Choi_Asilomar} use multiple antennas with antenna locations based on wavelengths such that the receive antenna falls within a null point of the transmit antennas.  A similar work in \cite{2012_MIDU} removes the dependency on the wavelength by using a fixed phase shifter to guarantee that a null point is created at the receive antenna.  A novel antenna design in\cite{2010_Patch_Sweden} uses two dual-polarized patch antennas to isolate the transmit and receive streams.  

Active techniques exploit the knowledge of the self-interference signal to cancel it from the received signal.  Experiments on the WARP platform in \cite{2012_Melissa_TWC} used an extra transmit chain to generate an up-converted RF cancellation signal that was combined with the incoming signal at the receive antenna.  A recent work in \cite{2013_Stanford_Sigcomm} gives a circuit design that essentially samples the RF self-interference signal and uses sinc interpolation to generate the cancellation signal.  An implementation in \cite{2010_low_power} combined the passive multiple antenna null point technique with an off the shelf noise canceler chip to generate the cancellation signal.  

Despite these cancellation techniques, complete cancellation of the self-interference signal has not been achieved to date.  An additional area of work is attempting to identify any fundamental limitations in radio transceiver designs which prevent complete cancellation.  A theoretical model was presented in \cite{2012_Achal_Asilomar,2013_Achal_TVT_Phase} which was used to show that phase noise in the local oscillators in the transmit and receive radio chains is the key bottleneck.  That work was then later extended in \cite{2013_Ashu_Globecom} to show that phase noise induced intercarrier interference (ICI) is the major limiting factor in an OFDM system.  Results in \cite{2012_Aalto_Asilomar, 2012_MIMO_relay} show another bottleneck appears in the form of limited dynamic range at the analog-to-digital converter.  Further hardware limitations and their impact on the eigenvalue spread of spatial transmit and receive covariance matrices was shown in\cite{2012_Hardwar_limitations}.  

Complementing the mostly implementation based work above are numerous theoretical results analyzing the gains of full-duplex.  A technique proposed in \cite{2010_vitual_FD} uses rapid on-off division duplex to achieve full-duplex capability with half-duplex radios.  Results in \cite{2012_Achal_comsnets} quantify the need for synchronisity between two communicating full-duplex nodes.  Work in \cite{2013_CISS_Aalto} derives the rate regions for a MIMO system and looks at the tradeoff between using passive and active techniques.  And finally, full-duplex communication can yield additional gains in various other applications.  Cognitive radios can now simultaneously sense network activity while sending  their own data in \cite{2012_Ashu_rates}.  Cooperative relay networks in \cite{2013_Jingwen_TWC} can exploit a full-duplex side-channel to improve the data-carrying bidirectional link performance.  

In the above discussion, cancellation techniques were organized as either active or passive.  A further classification can be made based on where along the transceiver chain does the cancellation occur.  A majority of both the passive and active techniques described above are implemented in the analog RF stage.  This occurs for two main reasons.  First, in order to not saturate the Low Noise Amplifier (LNA) at the front end of the receiver, it is crucial to provide a significant level of cancellation of the analog RF signal.  Typical transmission powers in mobile devices can easily reach 20 dBm and commonly used LNAs saturate at input power levels greater than -25 dBm.  This means that a minimum of 45 dB of cancellation is required at the analog RF stage alone.  

The second reason that most passive and active techniques operate in the analog RF stage is the accessibility and ease in which connections can be made with the radio circuitry.  As mentioned above, most existing wireless systems use radios designed for half-duplex.  Passive antenna designs, either the MIMO null point schemes \cite{2012_MIDU} or the polarized patch antennas \cite{2010_Patch_Sweden}, can be configured to connect to almost any type of radio regardless of the internal circuitry of the radio.  Similarly for active techniques, an analog RF cancellation signal can be easily added to the receive radio with a simple power combiner without modifying any of the radio circuitry \cite{2012_Melissa_TWC}.  Furthermore, if the active design needs access to the transmitted RF signal, that signal can also be easily obtained with a power splitter at the transmit radio \cite{2013_Stanford_Sigcomm}.  

Because of these two reasons, a majority of the analog cancellation techniques to date occur in the RF stage.  This has resulted in the \emph{analog baseband} stage receiving very little attention in terms of its capabilities for self-interference cancellation.  The major drawback for implementing analog baseband cancellation is that modifications to the radio circuitry are required.  Namely, a cancellation signal needs to be added after the RF down-converter but before the analog-to-digital converter, a location in the circuit not usually accessible.  However, in pursuit of the yet elusive complete cancellation of the self-interference signal, an analog baseband cancellation stage could help close the gap.  

\subsection{Summary of Contributions}
Our work in this paper proposes an analog baseband self-interference cancellation technique that is intended to both follow and complement an initial analog RF cancellation stage.  While our focus will be in the baseband stage, as described above, a successful realization of full-duplex hinges heavily on the analog RF  stage.  We outline our contributions in each of the stages below. 

\emph{Analog RF -} Similar to the patch antenna design proposed in \cite{2011_patch_antenna} we use a four-layer patch antenna prototype built by the Rice Integrated Systems and Circuits Lab at Rice University \cite{RISC_WEB}.  We consider two different versions of the prototype, one passive and one active.  Each version is characterized by different isolation and bandwidth values.   A network analyzer was used to measure the reflection coefficients and isolation of the prototype.  Using the isolation measurements, two different realizations of the full-duplex self-interference RF channel model are formed and the analytical equivalent baseband model is derived.  

\emph{Analog Baseband -} Using the derived self-interference baseband channel model, we then develop our analog baseband cancellation.  Training symbols are use to form an estimate of the self-interference channel.  The estimate is then used with the data symbols to generate a cancellation signal that is ideally equal to the self-interference signal.  
\begin{figure}[htp] 
\begin{center}  
  \includegraphics[width=0.45\textwidth]{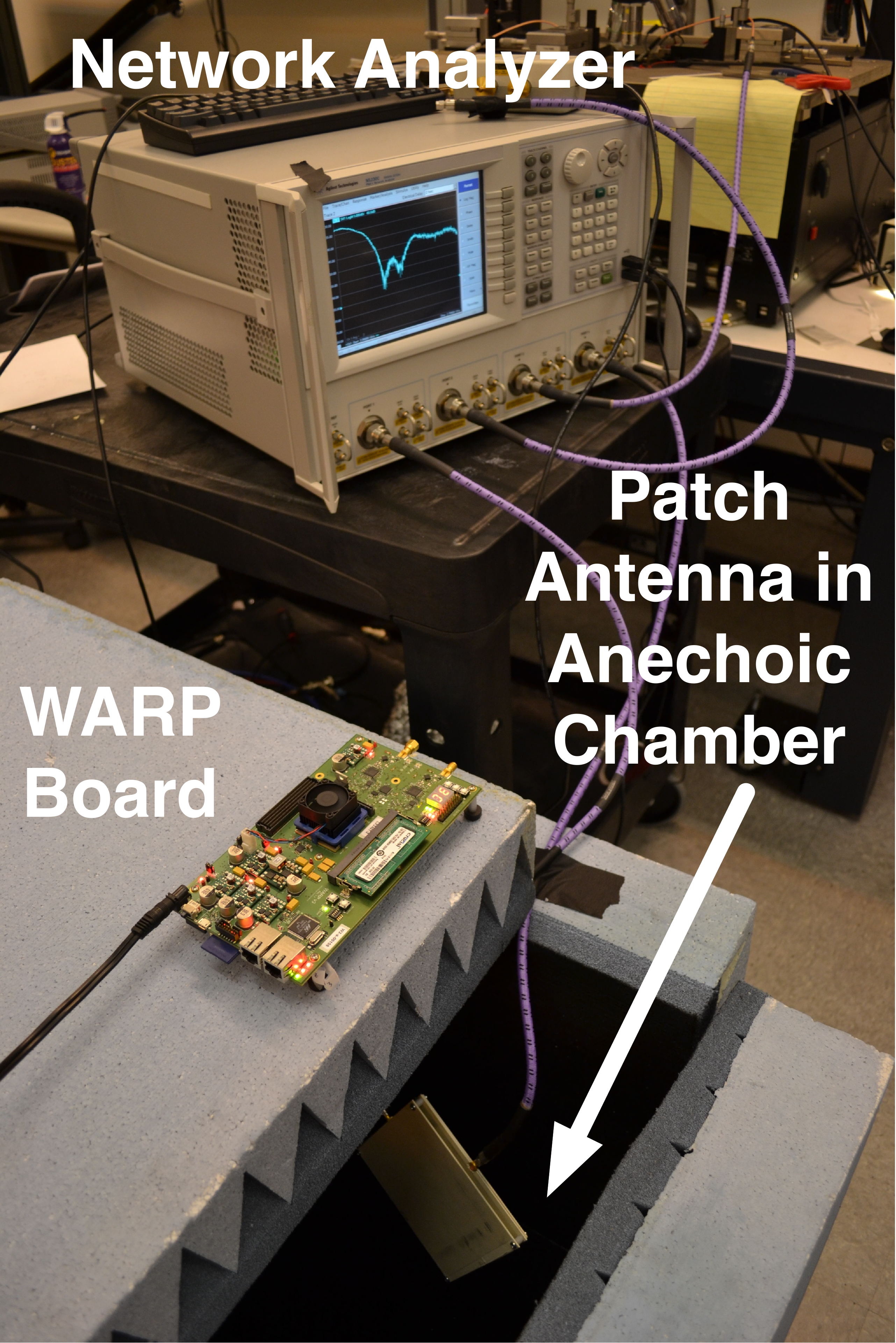} 
  \caption[fig:system]{Experimental setup with WARP board, full-duplex patch antenna prototype, and Network Analyzer.  }   
  \label{fig:setup} 
\end{center}  
\end{figure} 

\emph{Performance Evaluation -} We first evaluate the performance of the self-interference cancellation at a single user in terms of the quality of the desired signal relative to the self-interference signal.  We compare the performance of using only RF cancellation and a combined RF plus baseband cancellation.  Two key observations are made detailing different scenarios in which each cancellation scheme is more apt.  We then evaluate the performance of the self-interference cancellation from a communication link perspective between two full-duplex users.  We quantify the performance using the well known metrics of bit error rate probability and the achievable rate.  

The remainder of this paper is organized as follows.  In Section~\ref{sec:system_architecture} we define the system architecture which includes both the experimental setup and the analytical signal model.  In Section~\ref{sec:self_int_model} we provide details of how the self-interference is modeled.  Analysis of the self-interference cancellation technique appears in Section~\ref{sec:self_int_canc}.  A two-user full-duplex link is then evaluated in Section~\ref{sec:results} and concluding remarks are discussed in Section~\ref{sec:conclusion}.

\section{System Architecture}
\label{sec:system_architecture}
In this section, we introduce the experimental setup and provide the details for the corresponding functional block diagram and the analytical signal model of the system.  

\subsection{Experimental Setup}
We refer to Fig.~\ref{fig:setup} as we describe the experimental setup and highlight the three main components: the four-layer patch antenna, a network analyzer, and a WARP node.  The four-layer patch antenna prototype designed by \cite{RISC_WEB} contains both the transmit and receive antennas in a single form factor.  Both the transmit and receive antennas are fabricated on two-layer boards and then held in place parallel to each other with an air-gap between them.  As seen in the figure, the patch antenna was connected to an Agilent Network Analyzer and placed inside a small Anechoic chamber to remove the environmental effects on the measurements.  The network analyzer uses a  2.4 GHz high frequency test signal on the patch antenna and both isolation and phase measurements can be captured at the receiver.  The screen of the network analyzer in the figure shows an example realization of the isolation of the patch antenna.  

Also shown in the experimental setup in Fig.~\ref{fig:setup} is a WARP v3 node.  The WARP platform \cite{WARP_WEBSITE} is a scalable and fully programmable wireless testbed in which the patch antenna prototype was designed to be compatible with.  The WARP node implements a full OFDM wireless transceiver starting from the digital baseband, moving to the analog baseband, and finally passing to the analog RF.  The performance of the patch antenna as measured by the network analyzer will be the same when connected to the WARP node.  The WARP platform will then allow us to test baseband cancellation techniques in combination with the RF patch antenna prototype.


\subsection{Signal Model}
\label{subsec:signal_model}
We consider the two-node point-to-point wireless network shown in Fig.~\ref{fig:network}.  The two nodes, denoted as $a$ and $b$, are communicating with each other in a full-duplex manner in which the same temporal and frequency resources are being accessed simultaneously.  Each node has a single transmit antenna and a single receive antenna.  
\begin{figure}[htp] 
\begin{center} 
\includegraphics[width=0.42\textwidth]{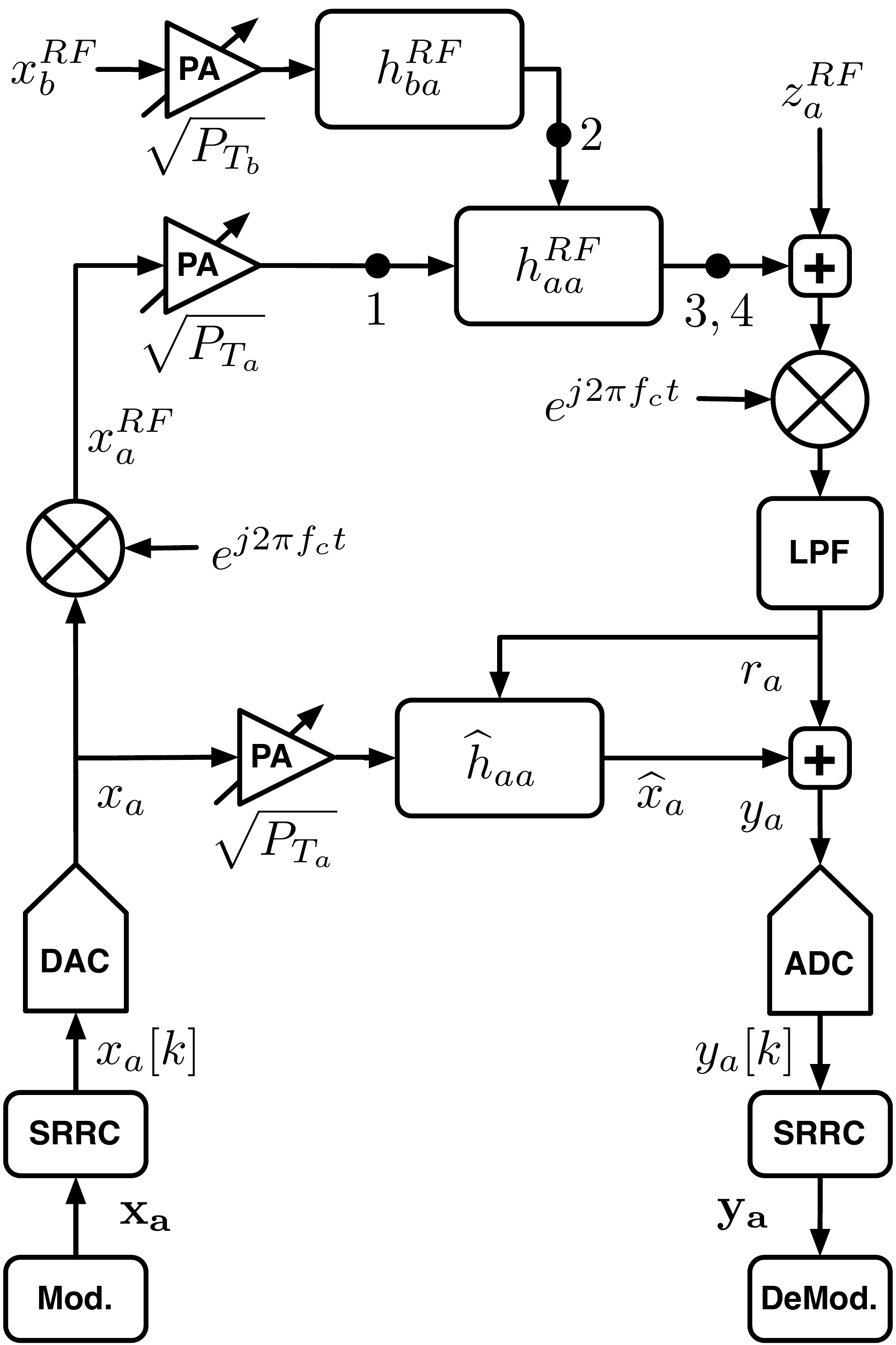}
  \caption[fig:system]{Functional block diagram of a full-duplex transceiver from the perspective of node $a$.  Terminal points ($\bullet$) 1, 2, 3, and 4 denote where different self-interference channels $h_{aa}^{RF}$ can be inserted.} 
  \label{fig:system}
\end{center} 
\end{figure} 
We refer to Fig.~\ref{fig:system} as we derive the signal model.  The functional block diagram shown in the figure illustrates the full-duplex transceiver considered in this network and is modeled after the transceiver design used on the WARP platform mentioned above. We note the signal model is derived from the perspective of node $a$ and that everything is identical for node $b$.  

At the transmitter side, $n_b$ bits in the $i^{th}$ bit vector $\mathbf{m}_i = [m_{i,1} \dots m_{i,n_b}]$ are first modulated (Mod.) into symbol $\mathbf{x_{a}}_{,i}$ of bandwidth $B$.  A signal constellation with $M$ points is used where $M = 2^{n_b}$.  We use $E[\cdot ]$ to denote the statistical expectation and assume $E[||\mathbf{x_{a}}_{,i}||^2] = 1$, which gives average unit energy over all symbols.The modulated symbols are then pulse-shaped using a square-root-raised-cosine filter (SRRC) which yields digital samples $x_a[k]$ which in turn serve as input to the Digital-to-Analog Converter (DAC).  Finally the baseband time domain signal $x_a(t)$ is outputted where we assume ideal DACs such that $x_a[k] \cong x_a(t)$.  For greater ease in the discussion, we remove the time notation $t$ and simply refer to $x_a(t)$ as $x_a$.  We will use this same notation for all other time domain signals that follow.  

The analog baseband signal is then up-converted to the carrier frequency $f_c$ which yields $x_a^{RF} = x_a e^{j 2\pi f_c t}$.  A power amplifier (PA) amplifies the signal with power $P_{T_a}$ and then the signal is transmitted by node $a$.  We note that we will use superscript $RF$ to denote the unconverted RF version of a signal.

After down-conversion and low pass filtering (LPF), the received baseband signal at node $a$ can be written as 
\begin{equation}
r_a = \sqrt{P_{T_b}}h_{ba}x_b + \sqrt{P_{T_a}}h_{aa}x_{a} + z_a,
\label{eq:r_bb}
\end{equation}
which is the sum of the self-interference signal $x_a$ through the self-interference channel $h_{aa}$, the signal of interest $x_b$ from node $b$ through the wireless channel $h_{ba}$, and additive white Gaussian noise $z_a$.  We will consider different realizations of the self-interference channel $h_{aa}$ and give more details in Section~\ref{sec:self_int_model}.  Finally, we assume that the noise is complex and Gaussian distributed as $z_a \sim \mathcal{CN}(0,\sigma_z^2)$.  We note that the system represented by (\ref{eq:r_bb}) is the baseband equivalent for the passband system at carrier frequency $f_c$.

Just prior to the Analog-to-Digital Converter (ADC), an analog baseband cancellation signal is added which gives 
\begin{equation}
y_a = r_a + \widehat{x}_a,
\label{eq:y_bb}
\end{equation}
as the time domain signal input to the ADC at node $a$.  Digital samples $y_a[k]$ are outputted from the ADC, and after the SRRC on the receiver side, the received symbols $\mathbf{y_{a}}_{,i}$ are determined.  Finally after demodulation of $\mathbf{y_{a}}_{,i}$, the $i^{th}$ bit vector $\mathbf{\widehat{m}}_i = [\widehat{m}_{i,1} \dots \widehat{m}_{i,n_b}]$ is received.  

As a final comment, we note that there is 
a sequence of mappings that map modulated symbols into digital samples, then to a baseband analog signal, and finally to a RF analog signal.  Based on this we know that the set $\{\mathbf{x_a}, x_a[k], x_a, x_a^{RF}\}$, and similarly $\{\mathbf{y_a}, y_a[k], y_a, y_a^{RF}\}$, represents the same information at different points on the transmit and receive chains.  We make this clarification because the focus of our work will be at the analog baseband stage, but there is still an interdependence between the various stages.   

\section{Self-Interference Model}
\label{sec:self_int_model}
We now give more details into how we model the RF self-interference channel $h^{RF}_{aa}$ and the corresponding baseband equivalent $h_{aa}$.  As mentioned above, the RF self-interference channel is modeled from real-time over-the-air measurements of a 
\begin{figure}[htp]
\begin{center} 
  \includegraphics[width=0.42\textwidth]{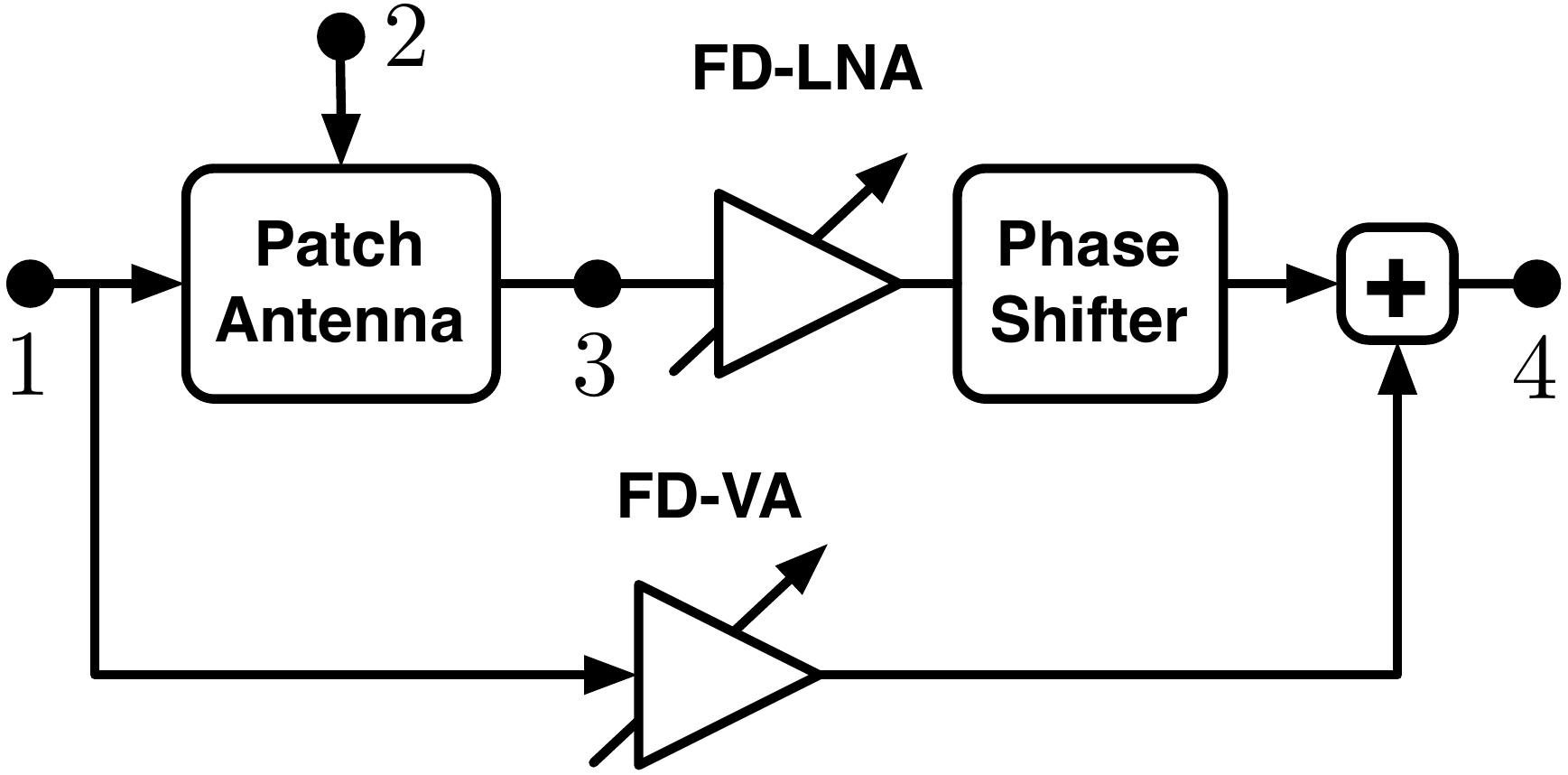}
  \caption[fig:system]{Block diagram for the full-duplex self-interference channel $h_{aa}^{RF}$ with terminal points ($\bullet$) 1, 2, 3, and 4 showing connection into the main block diagram.  Terminal points 1, 2, and 3 represent the \emph{Passive Suppression} scheme (PS).  Terminal points 1, 2, and 4 represent the \emph{Active Cancellation} scheme (AC).}
  \label{fig:RF_canceller}
\end{center} 
\end{figure} 
four-layer full-duplex patch antenna system prototype provided by \cite{RISC_WEB}. 
The block diagram for the antenna system prototype can be see in Fig.~\ref{fig:RF_canceller}.  We will consider two different variations of the block diagram and use the terminal points ($\bullet$) 1, 2, 3, and 4 in Fig.~\ref{fig:RF_canceller}, and their corresponding connections in Fig.~\ref{fig:system}, to distinguish between them.  
 
\subsection{Passive RF Suppression}
We refer to the part of the block diagram in Fig.~\ref{fig:RF_canceller} with terminal points 1 and 2 as input and terminal point 3 used as output as the passive cancellation (PS) scheme.  It consists of only the passive four-layer full-duplex patch antenna.  We denote the squared magnitude of the PS system in the passband by $|H^{RF}_{+}(f)|^2$, shown in Fig.~\ref{fig:chan_mag}, and the phase of the PS system by $\measuredangle H^{RF}_{+}(f)$, shown in Fig.~\ref{fig:chan_ang}. A maximum isolation value of 53.9 dB is obtained at a frequency of $f = $ 2.438 GHz.  Furthermore, it gives 42.5 dB of isolation across a 10 MHz bandwidth.  The phase of the PS system is approximately linear and decreasing with increasing frequency in the bandwidth of interest. 

\subsection{Active RF Cancellation}
We refer to the block diagram in Fig.~\ref{fig:RF_canceller} with terminal points 1 and 2 as input and terminal point 4 used as output as the active cancellation (AC) scheme.  The four-layer patch antenna is the first component in the circuit followed by a full-duplex low noise amplifier (FD-LNA) to amplify the incoming signal of interest.  After amplification, there is a phase shifter to undo some of the phase effects of the patch antenna.  There is a parallel path with a full-duplex variable amplifier (FD-VA).  The self-interference signal $x_a^{RF}$ will traverse both parallel paths.  The goal of the phase shifter and the FD-VA is to create two copies of the self-interference signal, one the inverse of the other.  The presence of the signal of interest in the top path prevents a perfect inverse from being formed and thus there will be some residual self-interference mixed with the signal of interest.  We note that this scheme uses a combination of passive and active techniques for cancellation, but denote it as the active AC scheme to clearly distinguish it from the purely passive PS scheme.  
\begin{figure}[htp]
\begin{center} 
  \includegraphics[width=0.45\textwidth]{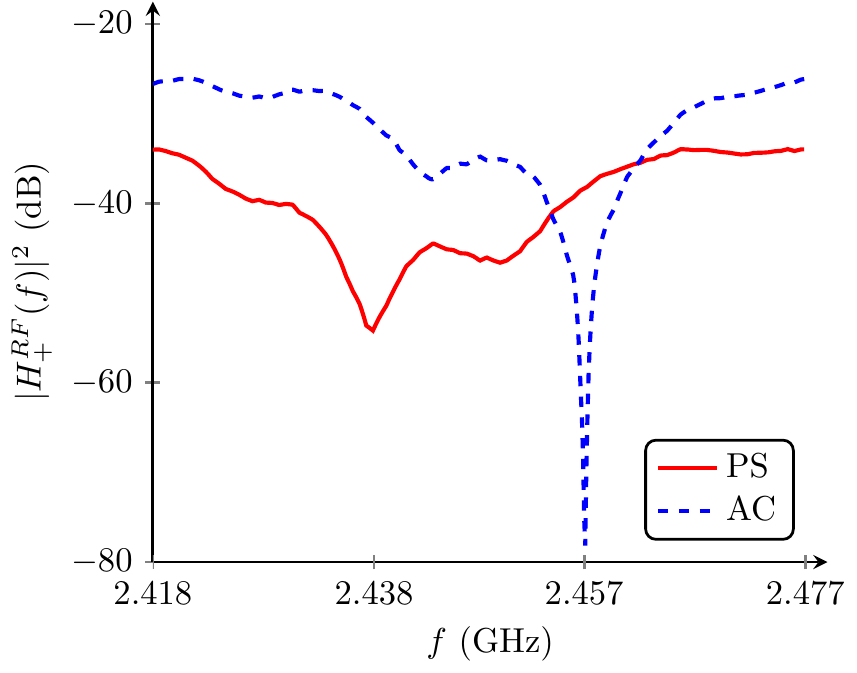}
  \caption[fig:chan_mag]{Isolation measurements for the full-duplex patch antenna system prototype.  The PS curve denotes the channel with only the passive patch antenna.  The AC curve denotes the channel with the combined passive and active elements.} 
  \label{fig:chan_mag} 
\end{center} 
\end{figure} 

\begin{figure}[htp]
\begin{center} 
  \includegraphics[width = 0.45\textwidth]{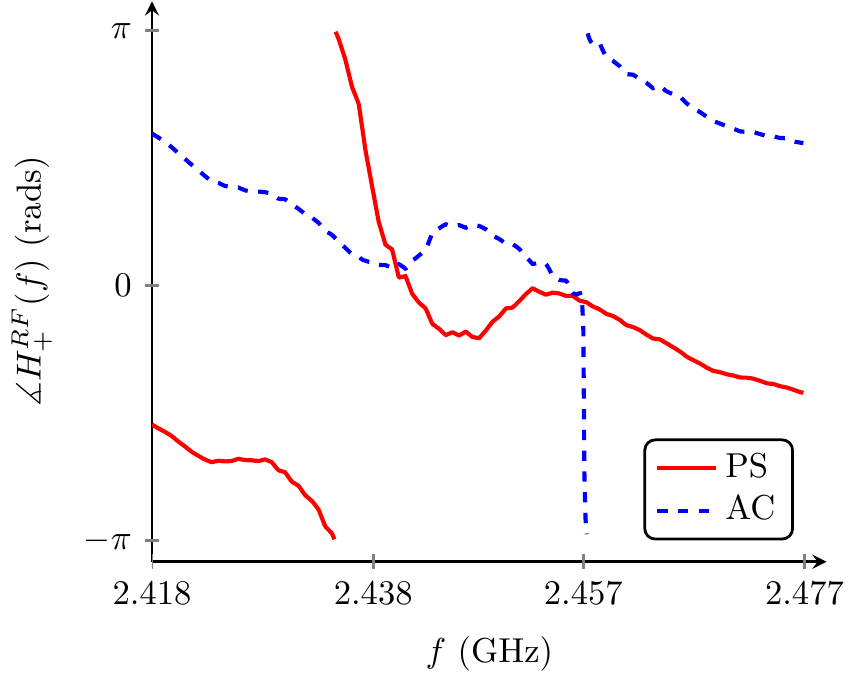}
  \caption[fig:chan_ang]{Phase measurements for the full-duplex patch antenna system prototype.  The PS curve denotes the channel with only the passive patch antenna.  The AC curve denotes the channel with the combined passive and active elements.} 
  \label{fig:chan_ang}
\end{center} 
\end{figure}

As was the case for the PS scheme above, both the squared magnitude and the phase of the AC scheme can be seen in Fig.~\ref{fig:chan_mag} and Fig.~\ref{fig:chan_ang} respectively.  The AC scheme provides the highest isolation of 78.1 dB at a frequency of $f = $ 2.457 GHz.  However, the high level of isolation is only achieved for a narrow band.  Only 35.3 dB of isolation is possible across a 10 MHz bandwidth.  Similar to the PS scheme, the phase of the AC system is approximately linear and decreasing with increasing frequency in the bandwidth of interest.  

\subsection{Equivalent Baseband Model}
We now derive the equivalent baseband model for the RF self-interference channel.  Each of the RF cancellation scheme prototypes give the maximum isolation at a different frequency.  As these designs are optimized for final versions, the circuits could be tuned to align each of their frequencies corresponding to their highest isolation.  However until those prototype designs are optimized, for analytical convenience, we will set the carrier frequency $f_c$ of our system to the optimal frequency of whichever RF cancellation scheme is used.  This assumption does not simplify any of the analysis nor does it affect the results.  The separation between the PS frequency of 2.438 GHz and the AC frequency of 2.457 GHz is just 19 MHz.  Changing the center frequency $f_c$ between these two values is equivalent to changing to a different 802.11 Wi-Fi channel.  

With those details in mind, measurements were made over a bandwidth $B_H$ centered at each corresponding center frequency $f_c$.  Using both the isolation and phase measurements, we can form
\begin{equation}
H^{RF}_{+}(f) =
\left\{ \begin{array}{cl}
|H^{RF}_{+}(f)|e^{j\measuredangle H^{RF}_{+}(f)},  & |f - f_c| \leq \dfrac{B_H}{2}\\
0, & \textrm{elsewhere}
\end{array}\right.
\end{equation}
which is the one-sided FFT of the passband channel.  Using properties of the Fourier transform, we can write
\begin{equation}
H_{aa}(f) = \dfrac{1}{2}H^{RF}_{+}(f+f_c),
\label{eq:H_si_f}
\end{equation}
which is the FFT of the equivalent baseband channel centered at 0 Hz.  The time domain representation of the self-interference channel can finally be written as
\begin{equation}
h_{aa} = \mathcal{F}^{-1}\{H_{aa}(f) \},
\label{eq:h_si}
\end{equation}
after taking the IFFT.  We note that the particular RF scheme used determines the specific realization of $h_{aa}$ for the baseband signal model.

\section{Analog Baseband Cancellation}
\label{sec:self_int_canc}
In this section we provide details on the analog baseband cancellation.  As above, we give the analysis and discussion from the perspective of node $a$.  
\subsection{Channel Estimation}
We utilize the Least Squares channel estimation to form a channel estimate of the self-interference channel $h_{aa}$.  While node $b$ remains silent, node $a$ performs the training by sending $N_{tr}$ training symbols.  Using (\ref{eq:r_bb}), we can write the received baseband signal in the training phase as
\begin{equation}
r_{a,tr} = \sqrt{P_{T_a}}h_{aa} x_{a,tr} + z_a,
\label{eq:r_train}
\end{equation}
where $x_{a,tr}$ and $r_{a,tr}$ are the transmitted and received training signals respectively.  Node $a$ can form an estimate of the channel by 
\begin{equation}
\widehat{h}_{aa} = \dfrac{r_{a,tr}x_{a,tr}^{-1}}{\sqrt{P_{T_a}}} = h_{aa} + \dfrac{z_a x_{a,tr}^{-1}}{\sqrt{P_{T_a}}},
\end{equation}
where we can see that the estimate $\widehat{h}_{aa}$ is the true channel corrupted by scaled additive noise.  Using the channel estimate, node $a$ can form the cancellation signal of $\widehat{x}_a = -\sqrt{P_{T_a}} \: \widehat{h}_{aa} x_a$, where $x_a$ is the data signal to be sent to node $b$ and the self-interference signal seen at node $a$'s receiver. 

\subsection{Cancellation}
Using (\ref{eq:y_bb}) and the cancellation signal just found, we can write
\begin{equation}
y_a = \sqrt{P_{T_b}}h_{ba}x_b + \sqrt{P_{T_a}}(h_{aa} -\widehat{h}_{aa}) x_{a} + z_a, 
\label{eq:y_a}
\end{equation}
which is the received analog baseband signal at node $a$ after cancellation.  We define the unwanted residual self-interference signal at node $a$ as
\begin{equation}
y_{a,res} \triangleq \sqrt{P_{T_a}}(h_{aa} -\widehat{h}_{aa}) x_{a} + z_{a},
\label{eq:y_a_res}
\end{equation}
and notice that the strength of the residual increases proportionally with channel estimation error.  Node $a$ needs to minimize the residual self-interference in an attempt to correctly receive the desired signal $h_{ba}x_b$.  Note that the additive noise is included in the residual self-interference signal.  This is an interference limited system so the addition of the additive noise is non-consequential to the self-interference.  
\begin{table}[h]
\begin{center}  
\caption{Network Simulation Parameters} 
\centering 
\begin{tabular}{|c||c|} 
\hline   
\textbf{System Parameters} & \textbf{Value} \\   
\hline\hline 
Bits per Symbol ($n_b$)& 2\\
\hline
PSK Modulation Order ($M = 2^{n_b}$)& 4 \\
\hline
Number of Data Bits ($N_{bits}$) & 2000\\ 
\hline
Number of Training Symbols ($N_{tr}$)& $5$\\ 
\hline
Carrier Frequency ($f_c$)& \{2.438, 2.457\} GHz  \\
\hline
Sampling Frequency ($F_s$)& 20 MHz  \\
\hline
Channel Bandwidth ($B_H$)& 20 MHz  \\
\hline
Signal Bandwidth ($B$)& 10 MHz\\
\hline
Node $a$'s Transmit Power ($P_{T_a}$)& 0 dBm\\
\hline
Received Power from Node $b$ ($P_{R_b}$)& -60 dBm\\
\hline 
\end{tabular} 
\label{table:vars}  
\end{center}
\end{table} 

The received analog baseband signal in (\ref{eq:y_a}) applies for either choice of $h_{aa}$ corresponding to the passive (PS) and active (AC) RF schemes.  We will use the label PS+B to denote when the analog baseband cancellation is used together with the passive RF scheme.  Similarly, we will use the label AC+B to denote when the analog baseband cancellation is used together with the active RF scheme.  

\subsection{Results}
Having just defined how the baseband cancellation works, we now quantify it's performance.  One of the more applicable performance metrics for this scenario is the \emph{Signal-to-Interference-Noise} (SINR) ratio.  If we look at the SINR at node $a$
\begin{equation}
\Gamma^a = \dfrac{E[|h_{ba}x_b|^2]}{E[|y_{a,res}|^2]},
\end{equation}
we can see that it relates the strength of the desired signal at node $a$ to the strength of the undesired residual self-interference.  As we will see in the results below, the strength of the desired signal from node $b$ is a key parameter in evaluating the performance of the analog baseband cancellation.  With that in mind, we define $P_{R_b} \triangleq E[|h_{ba}x_b|^2]$ as the received power of the desired signal.

\begin{figure}[htp]
\begin{center} 
  \includegraphics[width = 0.49\textwidth]{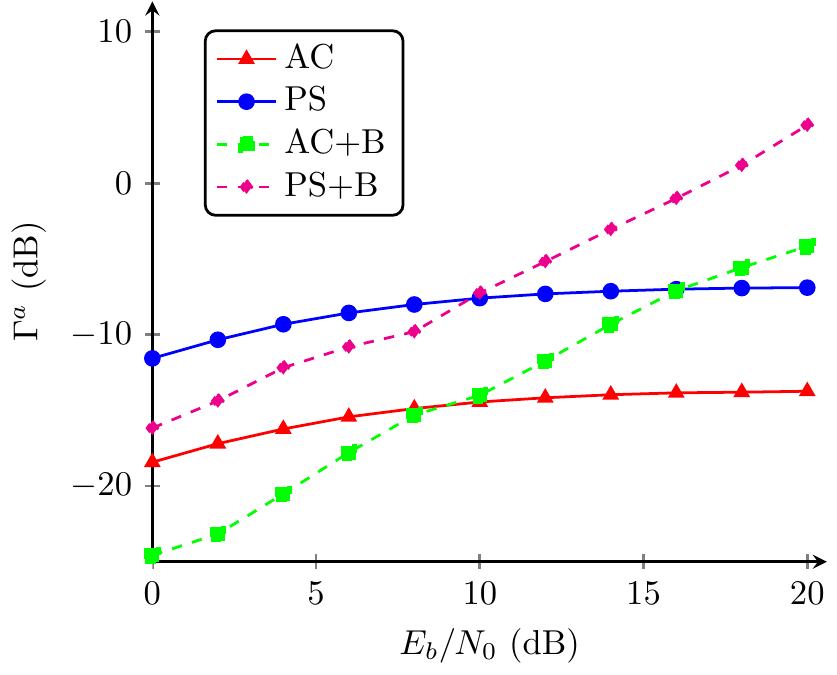}
  \caption[fig:chan_ang]{The signal-to-interference-noise ratio ($\Gamma^a$) of the desired signal from node $b$ to the residual self-interference at node $a$.  Baseband cancellation PS+B and AC+B improves the SINR as compared to the RF only PS and AC schemes.} 
  \label{fig:SIR_add_B}
\end{center} 
\end{figure}

The performance of the various cancellation schemes at node $a$ was simulated in Matlab using the parameters in Table~\ref{table:vars}. The parameter values shown in the table denote the default values unless explicitly specified for a particular figure  In Fig.~\ref{fig:SIR_add_B} the SINR is plotted versus $E_b/N_0$ to show the benefit of adding baseband cancellation to the RF only schemes.  At approximately $E_b/N_0 = 10$ dB, the SINR for both the PS and AC schemes begins to saturate while the SINR for the baseband PS+B and AC+B schemes continues to increase linearly with $E_b/N_0$.  We note that the value of $E_b/N_0$ is the same for both the transmitted data signals $x_a$ and $x_b$.  It is interesting to note that the passive RF scheme PS achieves a higher SINR as compared to the active RF scheme AC.  Similarly, the baseband scheme PS+B also achieves a higher SINR than the AC+B baseband scheme.  
Recall from the discussion above that the AC scheme provides a much higher level of isolation compared to the PS scheme but at a much smaller bandwidth.  

\begin{figure}[htp]
\begin{center} 
  \includegraphics[width = 0.49\textwidth]{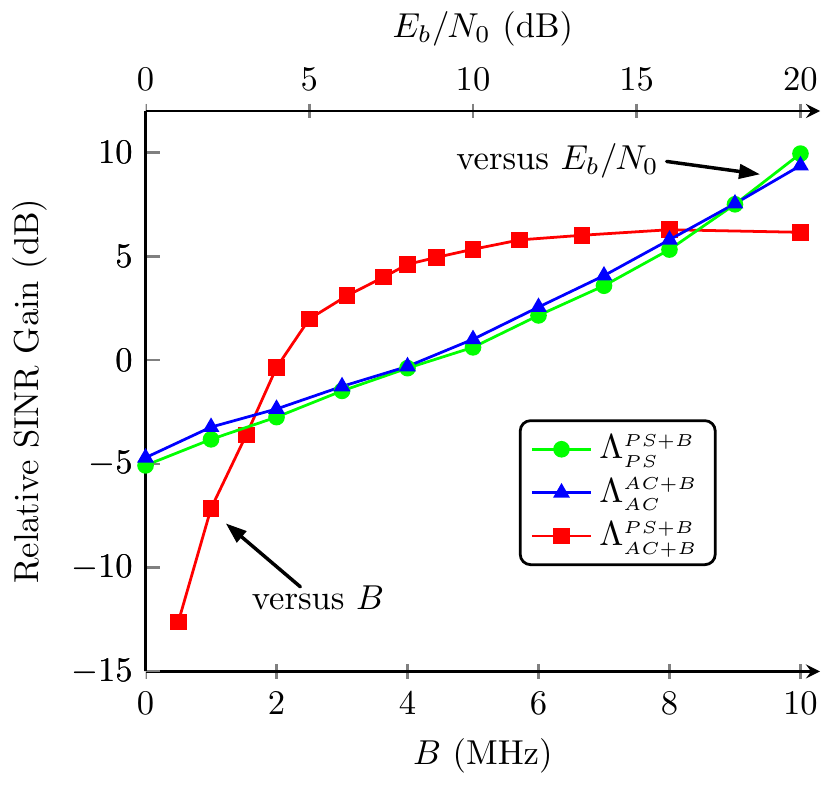}
  \caption[fig:chan_ang]{The relative SINR gain of the various cancellation schemes versus the bandwidth ($B$) and versus $E_b/N_0$.} 
  \label{fig:SIR_GAIN}
\end{center} 
\end{figure}

To quantify the tradeoff between the different scheme's isolation values versus bandwidth, we define the ratios
\begin{equation}
\Lambda^{PS+B}_{AC+B} = \dfrac{\Gamma^a_{PS+B}}{\Gamma^a_{AC+B}},
\label{eq:gamma_gain}
\end{equation}
\begin{equation}
\Lambda^{PS+B}_{PS} = \dfrac{\Gamma^a_{PS+B}}{\Gamma^a_{PS}},
\end{equation}
\begin{equation}
\Lambda^{AC+B} _{AC}= \dfrac{\Gamma^a_{AC+B}}{\Gamma^a_{AC}},
\end{equation}
to denote the relative SINR gains at node $a$ of the various schemes.  In Fig.~\ref{fig:SIR_GAIN}, the SINR gain of the PS+B scheme over the AC+B scheme, $\Lambda^{PS+B}_{AC+B}$, is plotted versus the bandwidth $B$.  Immediately we can see zero gain for a 2 MHz bandwidth meaning both baseband schemes achieve approximately the same SINR.  For a 10 MHz bandwidth, the PS+B scheme has about 8 dB higher SINR.  When nodes $a$ and $b$ transmit with a narrow bandwidth of 500 kHz, the baseband PS+B scheme actually performs approximately 13 dB worse than the AC+B scheme.  Also in Fig.~\ref{fig:SIR_GAIN}, the SINR gains of the baseband schemes compared to their RF only equivalents $\Lambda^{PS+B}_{PS}$ and $\Lambda^{AC+B} _{AC}$ are plotted versus $E_b/N_0$.  We can immediately see that the SINR gain is linearly increasing with $E_b/N_0$ when analog baseband cancelation is added to the RF only cancellation.  

\subsection{Main Observations}
Based on the above characterization of the cancellation schemes at node $a$, two important observations can be made.  First, adding baseband self-interference cancellation with PS+B or AC+B can significantly improve the quality of the desired signal at $a$ with respect to the self-interference signal at $a$.  The RF only schemes of PS and AC show marginal improvement in the SINR of the desired signal at $a$ for low $E_b/N_0$ values with diminishing effects at higher $E_b/N_0$.  

Second, the specific baseband cancellation scheme to use depends heavily on the bandwidth of the transmitted signals.  The high isolation, narrow bandwidth RF active cancellation scheme combined with baseband cancellation significantly outperformed the baseband PS+B scheme for a 500 kHz bandwidth.  But for bandwidths 2 MHz and larger, the PS+B scheme will show positive gains over the AC+B scheme.  In the rest of the analysis of this work, we will consider $B = 10$ MHz  as higher bandwidth systems are in more demand, and expect to see the PS+B scheme continue to outperform the AC+B scheme.


\section{Link Evaluation and Discussion}
\label{sec:results}
In this section we quantify the performance of the full-duplex link between nodes $a$ and $b$.  We first define the performance metrics and then we analyze the performance of the link under different network conditions.  

\subsection{Metrics}
The primary metric we use to quantify the link performance is the \emph{Probability of Bit Error} (BER).  We define the error probability as
\begin{equation}
P_b = \dfrac{1}{N_{bits}}\sum_{i=1}^{N_{bits}/n_b} \sum_{j=1}^{n_b} \mathbbm{1}(m_{i,j} \neq \widehat{m}_{i,j}),
\label{eq:ber}
\end{equation}
where the indicator function $\mathbbm{1}(m_{i,j} \neq \widehat{m}_{i,j})$ is used to denote when the $j^{th}$ received bit $\widehat{m}_{i,j}$ does not equal the corresponding transmitted bit $m_{i,j}$.  The bit errors are first counted over all $n_b$ bits in the $i^{th}$ symbol and then over all $N_{bits}/n_b$ symbols and finally averaged by the total number of bits sent $N_{bits}$.  

The second metric we consider is the achievable rate of the link between nodes $a$ and $b$.  We use the classical Shannon information theoretic notion \cite{cover_thomas} of the achievable rate to write $R^a= \log_2 (1+\Gamma^a)$ where the rate is measured in units of bits per second per Hertz (bps/Hz).  Many of our results thus far have quantified the performance of the baseband PS+B as compared to the AC+B scheme.  With that trend in mind, we quantify the rate difference between the achievable rate of the two different schemes as
\begin{equation}
\Delta R = R^a_{PS+B} - R^a_{AC+B},
\label{eq:rate_diff}
\end{equation}
where a positive difference denotes the performance gain of the PS+B scheme over the AC+B scheme.  

\subsection{Results}
\begin{figure}[htp] 
\begin{center} 
  \includegraphics[width = 0.49\textwidth]{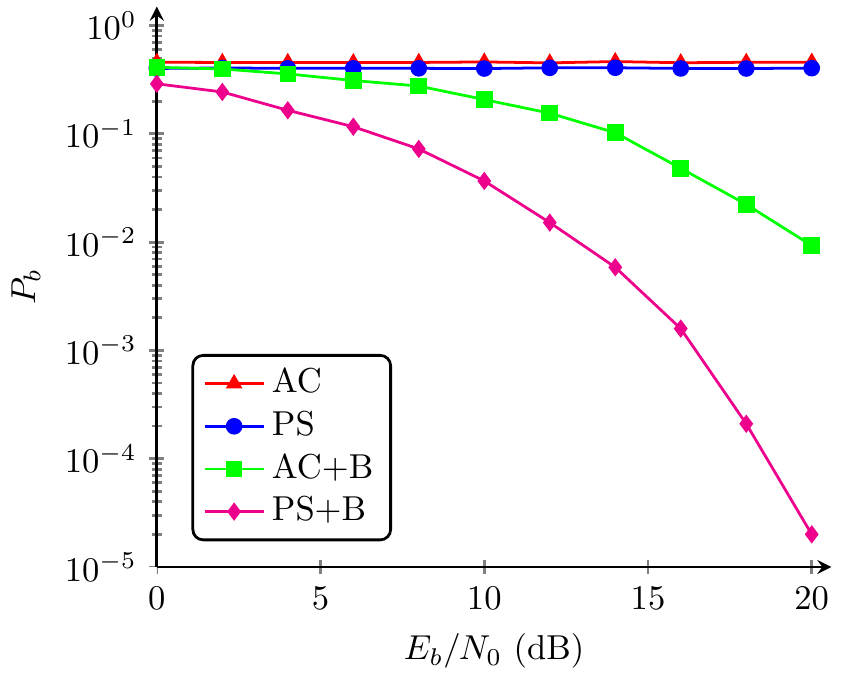}
  \caption[fig:chan_ang]{Probability of Bit Error ($P_b$) at node $a$ versus $E_b/N_0$.  The analog baseband cancellation in PS+B and AC+B significantly reduces the error rate as compared to the RF only cancellation schemes.} 
  \label{fig:BER_EbNo}
\end{center} 
\end{figure}

We now evaluate the link between node $a$ and node $b$.  We refer to the network parameters listed in Table~\ref{table:vars}.  We first look at the effect of of noise on the link.  Fig.~\ref{fig:BER_EbNo} plots the bit error probability $P_b$ with respect to $E_b/N_0$.  It is clearly noticeable how the use of baseband cancellation improves the link quality.   
For $E_b/N_0 = 10$ dB, we see a factor of 10 improvement in the BER for the PS+B scheme and only minor improvement for the AC+B scheme. At $E_b/N_0 = 20$ dB, up to $10^4$ improvement is observed.  
  
\begin{figure}[htp]  
\begin{center} 
  \includegraphics[width = 0.49\textwidth]{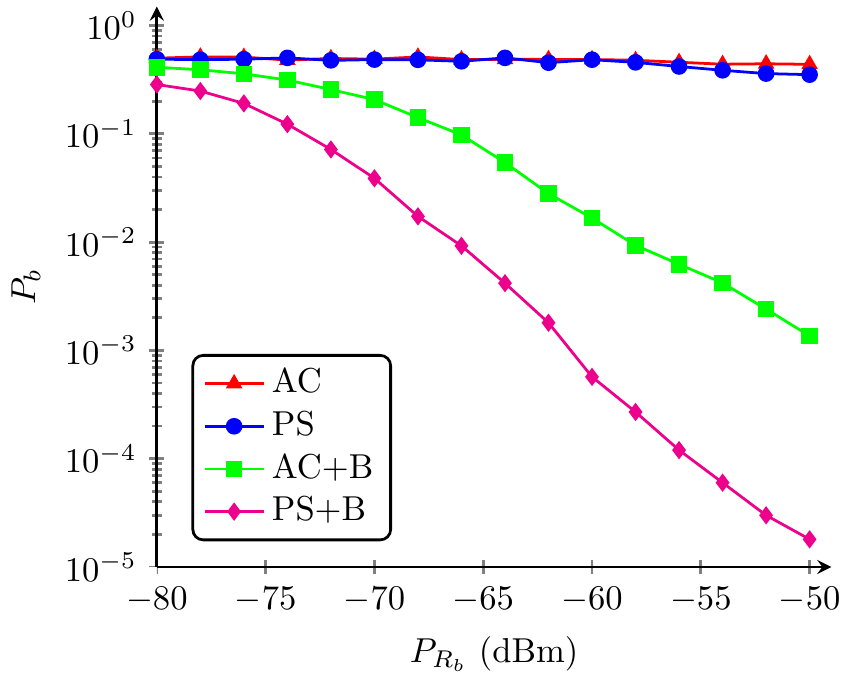}
  \caption[fig:chan_ang]{Probability of Bit Error ($P_b$) at node $a$ versus the received signal strength ($P_{R_b}$). The error rate decreases inversely proportional with the received signal power from node $b$.}  
  \label{fig:BER_Prb}
\end{center} 
\end{figure} 
We now look at the effect of the signal strength $P_{R_b}$ of the received signal from node $b$.  Fig.~\ref{fig:BER_Prb} plots the bit error probability with respect to $P_{R_b}$.  We can immediately see that the baseband cancellation improves the link reliability as the strength of the received signal increases.  A factor of $10^2$ improvement is realized for a signal strength of -63 dBm and a $10^4$ improvement is seen at a signal strength of -50 dBm.  We note that for large signal strength values, the RF only cancellation schemes AC and PS begin to show a slight improvement in the bit error probability.

\begin{figure}[htp] 
\begin{center} 
  \includegraphics[width = 0.49\textwidth]{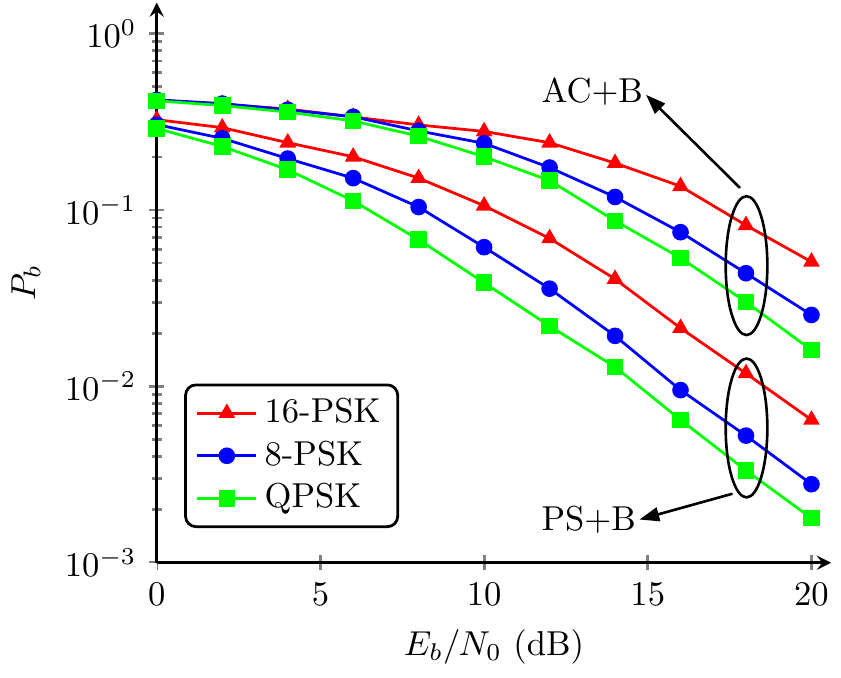}
  \caption[fig:BER_MPSK]{Probability of Bit Error ($P_b$) at node $a$ versus $E_b/N_0$ for different order PSK modulations.  The baseband PS+B scheme uniformly beats the AC+B scheme, but both schemes perform similarly with respect to the modulation order.  }  
  \label{fig:BER_MPSK} 
\end{center} 
\end{figure}

We now consider the effect of different order modulations on the bit error probability versus $E_b/N_0$.  Fig.~\ref{fig:BER_MPSK} shows the bit error probability $P_b$ for three different PSK modulation orders and compares the two baseband cancellation schemes PS+B and AC+B.  As is expected, the bit error rate increases with higher order modulation as the spacing in between data symbols decreases.  
A more surprising trend can be noticed in how close each of the different PSK curves are to each other.  In both the PS+B and AC+B sets of curves, the bit error rate for QPSK modulation is only separated from the 16-PSK curve by about 4 dB.  The same modulation orders are separated by 8 dB in an AWGN channel and about 9 dB for a Rayleigh fading channel, so the separation here is cut in half.  

\begin{figure}[htp]
\begin{center} 
  \includegraphics[width = 0.49\textwidth]{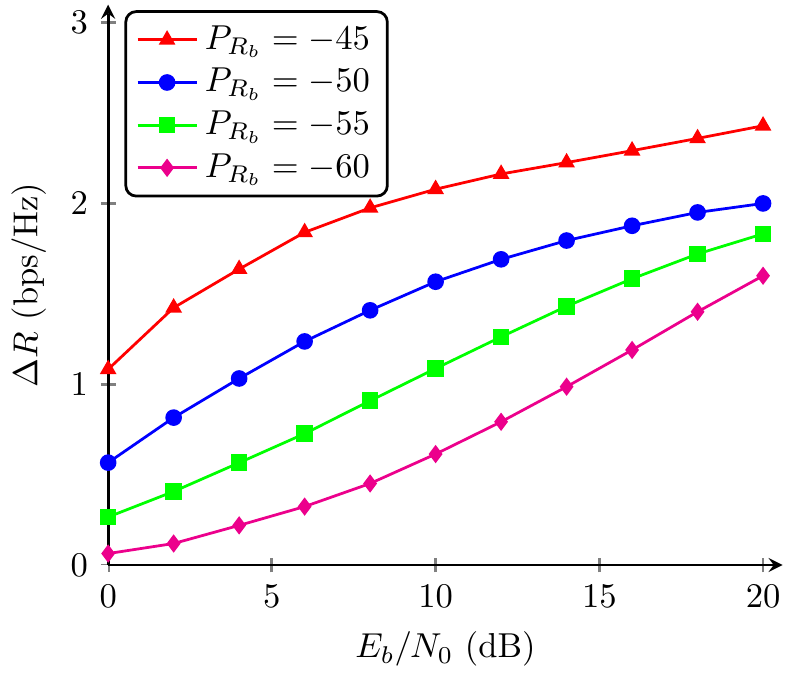}
  \caption[fig:rate]{The achievable rate difference ($\Delta R$) at node $a$ versus $E_b/N_0$.  The baseband PS+B scheme achieves up to 2.4 bps/Hz higher rate as compared to the AC+B scheme.} 
  \label{fig:rate} 
\end{center}  
\end{figure}

In Fig.~\ref{fig:rate}, the achievable rate difference from (\ref{eq:rate_diff}) for a link between node $a$ and node $b$ is shown for various values of the received signal power of node $b$'s signal at $a$.  As the strength of the desired signal increases, the gain of the PS+B scheme over the AC+B scheme increases.  Up to 2.4 bps/Hz improvement can be realized, but at high $E_b/N_0$, the rate difference begin to saturate and slowly converge.

\section{Conclusions}
\label{sec:conclusion}
We have proposed and evaluated an analog baseband self-interference cancellation scheme in this paper.  A prototype of a four-layer RF patch antenna was used in an experimental setup to provide real over-the-air measurements of a full-duplex self-interference channel.  Two variations of the prototype design, one passive and one active, were used to derive an analytical equivalent baseband model of the self-interference channel.  Channel estimation was used at the analog baseband stage to estimate the self-interference channel and create a cancellation signal.  

The performance of the analog baseband cancellation was quantified primarily with the signal-to-interference-noise ratio of the desired signal with respect to the residual self-interference signal.  It was observed that baseband cancellation can provide a linear increase in the SINR while RF only cancellation saturates to low SINR values.  A second notable observation is that at signal bandwidths above 2 MHz, the large isolation at narrow bandwidth provided by the active RF scheme becomes less effective.  At a bandwidth of 500 kHz, baseband cancellation combined with active RF can achieve 13 dB of gain, however at a larger 10 MHz bandwidth, baseband cancellation with passive RF achieves 8 dB of gain.  

Once the gains of analog baseband cancellation were established, the two-user full-duplex link quality was evaluated. First the the probability of bit error rate was used to quantify the performance.  It was observed that a $10^1 - 10^4$ reduction in BER was achieved by adding analog baseband cancellation to the RF cancellation schemes.  The highest gains were realized with the baseband PS+B scheme.  The full-duplex link was then evaluated in terms of achievable rate.  The rate difference between the two baseband schemes was calculated and the baseband PS+B scheme is able to achieve up to 2.4 bps/Hz improvement in achievable rate as compared to the AC+B scheme.  Based on these metrics, it is conclusive that a wide bandwidth, moderate isolation cancellation scheme is superior to a narrow bandwidth, high isolation scheme.  These key observations can help steer the design of future full-duplex radios.

\section{Acknoldgements}
The authors would like to thank the Rice Integrated Systems and Circuits (RISC) lab headed by Dr. Aydin Babakhani.  The patch antenna prototype was developed by graduate student researchers Tulong  Yang and Peiyu Chen.    
\ifCLASSOPTIONcaptionsoff
  \newpage
\fi



%
%
%

\bibliographystyle{IEEEtran} 
\bibliography{/Users/bkaufman/Dropbox/Brett/Brett_Papers/Bibs/full_duplex_brett,/Users/bkaufman/Dropbox/Brett/Brett_Papers/Bibs/Brett_Textbooks}   
\end{document}